\documentclass[12pt]{article}

\usepackage{newtxtext,newtxmath}

\usepackage{graphicx}

\usepackage[letterpaper,margin=1in]{geometry}

\renewenvironment{abstract}
	{\quotation}
	{\endquotation}

\date{}

\makeatletter
\renewcommand{\fnum@figure}{\textbf{Figure \thefigure}}
\renewcommand{\fnum@table}{\textbf{Table \thetable}}
\makeatother

\usepackage{scicite}

\usepackage{url}

\def\scititle{
    Physics-Aware Inverse Design for Nanowire Single-Photon Avalanche Detectors via Deep Learning
}
\title{\bfseries \boldmath \scititle}

\author{
        Boyang ~Zhang$^{1,2}$,
        Zhe ~Li$^{1\ast}$,
        Zhongju ~Wang$^{3}$,
        Yu ~Yang$^{1}$,\and
        H. Hoe ~Tan$^{1}$,
        Chennupati Jagadish$^{1}$,
        Daoyi ~Dong$^{2}$,
        Lan ~Fu$^{1\ast}$\and
	\small$^{1}$Australian Research Council Centre of Excellence for Transformative Meta-Optical Systems, \and \small Research School of Physics, The Australian National University, Acton, Canberra, 2601, Australia.\and
	\small$^{2}$School of Engineering, The Australian National University, Acton, Canberra, 2601, Australia.\and
    \small$^{3}$School of Systems and Computing, The University of New South Wales, Campbell, Canberra, 2612, Australia.\and
	\small$^\ast$Corresponding authors. Emails: zhe.li@anu.edu.au, lan.fu@anu.edu.au
}

\begin{document} 

\maketitle

\begin{abstract} \bfseries \boldmath
Single-photon avalanche detectors (SPADs) have enabled various applications in emerging photonic quantum information technologies in recent years. However, despite many efforts to improve SPAD's performance, the design of SPADs remained largely an iterative and time-consuming process where a designer makes educated guesses of a device structure based on empirical reasoning and solves the semiconductor drift-diffusion model for it. In contrast, the inverse problem, i.e., directly inferring a structure needed to achieve desired performance, which is of ultimate interest to designers, remains an unsolved problem. We propose a novel physics-aware inverse design workflow for SPADs using a deep learning model and demonstrate it with an example of finding the key parameters of semiconductor nanowires constituting the unit cell of an SPAD, given target photon detection efficiency. Our inverse design workflow is not restricted to the case demonstrated and can be applied to design conventional planar structure-based SPADs, photodetectors, and solar cells.
\end{abstract}


\subsection*{Introduction}
\noindent
Single-photon detector (SPD) technologies have a wide range of applications where very high sensitivity detection of light is essential, such as bioluminescence detection and imaging, time-of-flight 3D scanners, light detection and ranging (LiDAR), time-resolved photoluminescence, and more recently, the optical quantum information applications, including quantum encryption, long-distance free-space quantum communication, and photonic quantum computing \cite{Hadfield2009,Eisaman_invited_SPD_review,Ceccarelli_etal_SPD_QuantumApp_review,Weida_Hu_SPD_review_small}. A single-photon detector, in its simplest, ideal form, is a device that can generate one electrical output pulse in response to a single input photon. Two figures of merit are critical to achieving such high sensitivity: photon-detection efficiency (PDE) and dark count (DCR). PDE is the probability that a photon incident on the detector leads to an electrical output large enough to be registered by the external electronics, and DCR represents the average number of counts registered by a detector per second in the absence of any input photons. A perfect SPD would have a detection efficiency of 100\% and zero dark counts. 

Over the past two decades, significant efforts have been devoted to developing different SPD technologies aiming at increasing PDE and suppressing DCR \cite{Weida_Hu_SPD_review_small}. Among them, superconductivity-based detectors like superconducting nanowire and transition-edge detectors can achieve PDEs beyond 95\% \cite{Reddy_SNSPD_Optical_2020,Chang_SNSPD_APLPhotonics_2021} and low DCRs (10$^{-2}$ Hz) \cite{Zhang2011_SNSPD_LowDCR_2011,Wollman_SNSPD_lowDCR_OE_2017}. However, these characteristics have never been demonstrated simultaneously in the same device due to the technological trade-offs arising from the fabrication and operation of the device. Furthermore, a critical bottleneck of superconducting-based detectors is their requirement of cryogenic temperature ($<$4 K), which significantly adds to the cost and complexity of the experimental setup and becomes prohibitive to use for applications such as satellite optical receivers \cite{Ceccarelli_etal_SPD_QuantumApp_review}. Semiconductor-based single-photon avalanche detectors (SPADs), on the other hand, can operate at room temperature or slightly below. SPADs, mainly based on Si and InGaAs/InP materials, are arguably the dominant technology in today's commercial market thanks to their reliability, compactness, low operation voltage, competent detection efficiency in the visible and near-infrared (NIR) wavelengths, and ease of being integrated into a detector array and compatibility with other parts of peripheral CMOS logic circuits \cite{Eisaman_invited_SPD_review,Ceccarelli_etal_SPD_QuantumApp_review}. SPADs rely on the large avalanche current triggered by a physical process called impact ionization when absorbing a single photon. When an SPAD is biased beyond its breakdown voltage, known as operating in the Geiger mode, a photo-generated electron or hole can gain enough kinetic energy in such an elevated electric field and initiate a cascade of collisions with the host atoms, knocking additional electrons and holes out of lattice bonds. The chain reaction creates an avalanche of mobile electrons and holes constituting a measurable current. On the one hand, working under an extremely high electric field is desirable as it greatly increases the probability of a single-photon-induced electron or hole triggering a self-sustaining avalanche, leading to a high PDE. On the other hand, a large electric field also adversely increases the chance of thermally generated carriers becoming energetic, initiating impact ionization, and generating false output pulse indistinguishable from photon-induced output, giving a high DCR. Hence, an SPAD with high detection efficiency usually has a large dark count rate.

A major challenge to advancing SPAD technologies is, therefore, tackling the performance trade-off between PDE and DCR, both of which increase under elevated electric fields due to the avalanche process. To optimize the overall device performance, physics-based device modeling enabled by technology computer-aided design (TCAD) tools is normally employed to accelerate the design process and reduce the cost associated with physical prototypes, which are expensive to fabricate, especially when multiple iterations are necessary. Fig. \ref{fig:fig1_conventional_workflow} summarizes the conventional SPAD design workflow \cite{Li_SPAD_workflow_2023}. Here, we use SPADs made of indium phosphide (InP) nanowire array as an example shown in Fig. \ref{fig:fig1_conventional_workflow}a. Nanowire-based SPADs have recently drawn increasing research attention as they promise to achieve much higher PDE than their planar-structure counterparts, thanks to the extreme light-trapping effect enabled by the periodic array arrangement and dimension tunability \cite{Anttu_2019,Gibson2019,Vidur_ZhuYi_AM_2021}. Meanwhile, their nanoscale size and high material quality ensure a low DCR despite operating in a high-PDE condition, since DCR is directly proportional to the device volume and material's defect density \cite{Huffaker_NWSPAD_2019,Weida_Hu_SPD_review_small}. The conventional SPAD design paradigm starts with populating a limited range of material and structural properties as input parameters for the physics-based simulators. Then, it generates key intermediate variables, from which several SPAD performance metrics, e.g., a pair of PDE and DCR, at a given bias, device structure, and material properties can be calculated, as illustrated in Figs. \ref{fig:fig1_conventional_workflow}a to e. This process is also termed ``forward" calculations or designs. By sweeping key parameters of interest, which usually takes several tens of hours to days, depending on the experimental constraints and specific applications, a detailed mapping between PDE and DCR can be profiled and used for design considerations. An example of such mapping by varying the intrinsic region thickness, a key parameter determining the avalanche breakdown voltage, is shown in Fig. \ref{fig:fig1_conventional_workflow}e (lower panel), highlighting the typical SPAD's performance trade-off. The entire workflow must be performed iteratively to optimize the device structure until the design objective and desirable performance metrics are satisfied. This process is termed ``inverse" design. It is clear that the inverse design is of most interest to designers since the ultimate goal is to achieve a specific PDE or DCR, and the particular structure employed for this task may be less important.  

The key drawback of the conventional workflow in Fig. \ref{fig:fig1_conventional_workflow} is that it still relies on many cycles of empirical reasoning and trial-and-error modeling to achieve the desired performance, which is inefficient and often ineffective. This bottleneck arises due to two factors commonly present in semiconductor device development: a large design space and a relatively poor physical understanding of how each design parameter affects performance \cite{Glasmann_2021}. For SPAD design, in particular, the highly non-linear, indirect relationship between the input parameters, e.g., structural and material properties, and the target performance metrics like PDE and DCR, makes it increasingly challenging to understand parameter-performance correlations as the number of free parameters increases. This is further complicated by the design trade-off between PDE and DCR. 

To alleviate the design bottleneck, we consider a deep-learning-based approach. In recent years, deep learning (DL) techniques, especially the deep neural networks (DNNs), have emerged as a powerful computational tool in research areas beyond computer science \cite{Hinton_etal_2015} and have been successfully employed in physics \cite{Baldi_etal_2015,Raissi_JCP_Milestone_2019,Brunton_ML_Review_Fluid_Mech,ML_in_phys_2024}, chemistry \cite{Mater_and_Coote_ChemRev_2019,Bruno_etal_DL_in_AnalyChem_2021,Goh_etal_DL_in_CompChem_2017}, biology \cite{Zhou2015,Alipanahi2015,Jumper2021}, and material science \cite{Choudhary2022,Merchant2023}. In particular, DL techniques have found numerous novel applications in both forward and inverse designs of artificial electromagnetic materials (AEMs), including photonic crystals, nanoplasmonics, metasurfaces, and metamaterials \cite{Ma_NaturePhotonics_2021,Wiecha_Review_2021,ChenWei_AM_review_2024}. In those cases, the forward design, which would otherwise be performed by the time-consuming full-wave simulations based on the finite-element method (FEM) or finite-difference time-domain (FDTD) method, is empowered by DNN-based surrogate models to directly predict electromagnetic scattering responses given by a set of structure geometries with high speed and accuracy. The inverse design, though significantly more challenging, is under extensive research and has led to the development of a range of specially designed models based on generative adversarial networks, neural adjoint, and variational autoencoder methods \cite{Wellie_AEM_DL_Review_AFM2021,Willie_InvDesign_Benchmark_Nanoscale_2022,JayGuo_InvDesign_Benchmark_OES_2022,Li_Metasurface_inverse_design_review_2022}, to name a few. 

Inspired by the recent advancements in deep learning for the design of nanophotonic devices, we propose an inverse design workflow to shift the current SPAD design approach in at least two ways: 1) by significantly accelerating the evaluation of various candidate structures in forward designs, which replaces the most time-consuming step (Fig. \ref{fig:fig1_conventional_workflow}b) in a conventional workflow, and 2) by directly inferring the key design parameters needed to achieve target figures of merit of an SPAD, which replaces the convoluted iterative process in Fig. \ref{fig:fig1_conventional_workflow} and would be the ultimate interest to the research and industrial communities. More specifically, we identify two key intermediate variables for the forward design, the electric field within a device beyond avalanche breakdown and the quantum efficiency at zero bias. We construct two separate DNN-based surrogate models, connecting SPAD structural and electronic properties to these two responses, respectively. These two forward models were trained with high-fidelity data generated by physics-based semiconductor device simulations. For the first time, we demonstrate that artificial neural networks can be trained to accurately predict avalanche electric field and quantum efficiency based on a subset of an SPAD's architectural parameters. In this case, we include the nanowire lengths and doping concentrations of the p-, n-doped and intrinsic regions, as well as the applied voltage as input features for the neural networks. Subsequently, PDE and DCR can be obtained in the same way as in the conventional workflow (Fig. \ref{fig:fig1_conventional_workflow}e), which are calculated with fast postprocessing routines and are not a bottleneck of the entire workflow. The proposed surrogate model, once trained, can significantly accelerate the forward design by reducing the computation time from several hours with physics-based simulators to milliseconds. Based on the highly accurate forward surrogate model, we constructed an inverse design workflow that takes a target PDE and outputs the required region lengths, doping, and the applied bias of an SPAD. The accuracy of these inversely predicted parameters can be verified by feeding back into the physics-based device simulator, and we demonstrated that the average error is less than 10\%. The inverse design model also computes a range of DCRs for a desired PDE, informing the designers of the SPAD performance trade-off and facilitating the device design. Lastly, it should be noted that although we demonstrated the proposed inverse design workflow with InP nanowire-based SPADs, the approach itself is material and device-agnostic and can be applied to other semiconductor systems with modified material parameters and device structures. It provides a new paradigm in the design of optoelectronic devices, laying the foundation for the inverse design of photodetectors and solar cells. 

\subsection*{Results}\label{sec:results}

\subsubsection*{Designing forward prediction networks}\label{sec:results_forward_networks}
\noindent
A single-photon detection process involves two steps: first, an incident photon is absorbed by the detector, generating an electron-hole pair; then, the photo-generated electron-hole pair must trigger avalanche breakdown such that the detector produces a large enough current to register the signal. In terms of SPAD device modeling, these two steps are characterized by two intermediate quantities, quantum efficiency (Fig. \ref{fig:fig1_conventional_workflow}d) and electric field (Fig. \ref{fig:fig1_conventional_workflow}c). Quantum efficiency indicates how well an incident photon is absorbed, and the electric field is used to calculate the avalanche triggering probability, the chance that an electron-hole pair can trigger an avalanche event \cite{Li_SPAD_workflow_2023}. Combining quantum efficiency, avalanche triggering probability, and extra material properties (e.g., defect densities), the key SPAD performance metrics like photon detection efficiency (PDE) and dark count rate (DCR) can be readily obtained, as illustrated in Fig. \ref{fig:fig1_conventional_workflow}e.

A conventional SPAD design workflow predicts PDE and DCR by solving a set of three coupled partial differential equations, i.e., the drift-diffusion model, for electric potential ($\phi$), and electron ($n$) and hole ($p$) concentrations, as shown in Fig. \ref{fig:fig1_conventional_workflow}b. Obtaining a physically correct electric field is critical for computing avalanche triggering probability, thus giving an accurate PDE value (see more details in Methods). It has been shown that, in order to correctly recover the space-charge effect present in an electric field once avalanche breakdown occurs, it is necessary to solve the original, time-dependent drift-diffusion model (Fig. \ref{fig:fig1_conventional_workflow}c) \cite{Li_SPAD_workflow_2023}. This limits avalanche current from approaching infinity (which is unphysical) and avoids numerical instability and divergence. A key bottleneck is that solving time-dependent drift-diffusion models is very time-consuming since avalanche breakdown involves high non-linearity and exponential increase of carriers in a very short period. Typically, an SPAD simulation sweeping the applied voltage up to 5 V after avalanche breakdown can take five to ten hours, depending on the nanowire length and doping. Therefore, our first task is to replace the time-dependent drift-diffusion model with a fully connected feedforward deep neural network as a surrogate model that can capture the space-charge effect and accurately predict a spatial electric field distribution along the extracted one-dimensional nanowire SPAD structure under applied voltages beyond avalanche breakdown. Here, without loss of generality, we consider that nanowires have a p-i-n structure (Fig. \ref{fig:fig2_forward_network_illustration}a), the most common structure for SPADs, which can be routinely grown by the metal-organic chemical-vapor deposition (MOCVD) technique \cite{Ziyuan_IIIV_SingleNW_SolarCell_Review_2018,JennyWong_NW_review_AM_2020}. Generally, many structural and material-related parameters could affect the distribution of an avalanche electric field. Based on the semiconductor device physics, we narrow the number of input parameters down to the seven most essential and experimentally accessible ones: the p-/i-/n-region lengths ($L_p$, $L_i$, and $L_n$), the doping concentration of each region ($D_p$, $D_i$, and $D_n$), and the applied voltage ($V_a$). The region lengths and doping uniquely define a detector's built-in electric field profile, which affects where the peak photon absorption region is located and where a photo-generated electron-hole pair would accelerate to trigger an avalanche. The applied voltage directly controls the electric field strength within a detector, and the greater the field strength, the more the avalanche events. Therefore, a larger applied voltage leads to a higher PDE but also a larger DCR. The network architecture for electric fields is shown in Figs. \ref{fig:fig2_forward_network_illustration}a-c. It deals with the regression problem between key structural and electrical parameters (segment length, doping, and bias) and the complicated spatial distribution of the electric field along a nanowire.

Calculating quantum efficiency in a conventional SPAD workflow is a much faster routine. It involves solving the steady-state drift-diffusion model (Fig. \ref{fig:fig1_conventional_workflow}d), commonly used in modeling photodetectors and solar cells since it only characterizes photon absorption efficiencies regardless of whether there is an avalanche. We trained a separate neural network for quantum efficiency because the underpinning device physics and the physical processes to be captured are very different, as illustrated in Fig. \ref{fig:fig1_conventional_workflow}b (see more in Methods). For calculating avalanche electric field, the source term in the drift-diffusion equations accounts for impact ionization, whereas for quantum efficiency, the source term is the light-induced carrier generation rate. Doing so also increases the overall computational efficiency and facilitates easy integration into our proposed inverse design workflow. We also assume that quantum efficiency has a negligible dependence on the applied voltage so that the input for the corresponding network only takes the first six components as its input without the information on applied voltages. The network architecture is shown in Figs. \ref{fig:fig2_forward_network_illustration}a, b and d, with the same network depth and width as of the electric field's network. Its output is a spatial distribution of the quantum efficiency along a nanowire.

For a demonstration of our forward design workflow, we use an InP nanowire array-based SPAD operated at the temperature of 150 K as an example, with a typical detector structure shown in Fig. \ref{fig:fig1_conventional_workflow}a and an enlarged nanowire unit cell shown in Fig. \ref{fig:fig2_forward_network_illustration}a. The high-fidelity physics-based training data were generated by the Semiconductor Module, COMSOL Multiphysics software, using the conventional workflow depicted in Fig. \ref{fig:fig1_conventional_workflow}. The design space, as summarized in Fig. \ref{fig:fig2_forward_network_illustration}a, consists of vectors of seven device input parameters. Ideally, all the inputs would be best sampled independently and identically distributed for high training quality. Here, we consider the experimental constraints in the dimension and doping control during typical selective-area epitaxy MOCVD growth of InP nanowires. For example, the p-region is usually the substrate from which the nanowire is grown, whose length and doping concentration are always fixed. The i-region and n-region together constitute the actual nanowire, whose lengths can be controlled relatively easily by the state-of-the-art growth techniques \cite{Shiyu_AM_selfpowered_2022}, while fine doping manipulation remains a challenging issue so far \cite{JennyWong_NW_review_AM_2020}. Based on the experimental and theoretical studies \cite{QianGao_InP_NW_review_2019,Anttu_2019,JennyWong_NW_review_AM_2020,Gibson2019,Vidur_ZhuYi_AM_2021}, we adopt a subset of otherwise infinitely large and unrealistic design space: the p-region length ($L_p$) is fixed at 1 $\mu$m (as it usually represents a substrate), and the i-region length ($L_i$) is varied from 600 to 1800 nm with a step size of 50 nm, while the n-region length ($L_n$) is adjusted accordingly to keep the total length a constant of 3 $\mu$m. The p- and n-region doping is $D_p=D_n=10^{18}$ cm$^{-3}$, and the i-region is slightly n-doped ($D_i=5 \times 10^{15}$ cm$^{-3}$) due to unintentional background doping commonly observed in undoped InP nanowires grown by the selective-area MOCVD \cite{QianGao_InP_NW_review_2019}. The applied voltages used for the electric field network are the ones swept from the threshold breakdown voltage to 5 V above with a step size of 0.2 V. Generally, the threshold breakdown voltage is different for different combinations of region lengths and doping. Overall, the variation of the nanowire dimension gives 25 sets of different input configurations; combined with 26 sets of applied voltage points for each dimension variation, it gives 650 sets of input vectors. To increase the training data size, we also swap the input variable $L_n$ with $L_p$, and $D_n$ with $D_p$, and reverse the orders of output electric field and quantum efficiency data accordingly, which gives us, in total, 50 sets of training data for the quantum efficiency network and 1300 sets for the electric field network. The extra data sets greatly enhance the training quality, and they are independent of the original sets, thanks to the intrinsic structural symmetry and the fact that the neural network's output has no inherent preference for the device axis, i.e., the network should make good inferences regardless of the p- or n-region starting with $x=0$ of the 1D nanowire axis. 

The forward prediction network for electric fields has an output layer uniformly sampled to 1000 spatial grid points along the 1D nanowire axis from 0 to 3 $\mu$m, large enough to fully resolve the huge transition at the p-i and n-i junction interface where the avalanche electric field peaks (see Fig. \ref{fig:fig2_forward_network_illustration}c). The supervised training process is performed by minimizing the loss function, which measures the squared differences between the electric field prediction generated from the network and the high-fidelity physics-based simulation results. Among the 1300 sets of different configurations, 80\% are assigned to the training set, 10\% to the validation set, and the remaining 10\% to the testing set. Central to the network's accuracy is the ability to predict the peak avalanche electric field values around the p-i and n-i junction regions, typically one or two orders of magnitude higher than those in the bulk p- and n-regions. This simply manifests the space-charge effect based on the device physics. We find that even a minor difference between the predicted and true values by $\sim$10\% could result in a significantly underestimated avalanche-triggering probability, introducing a huge error in the final PDE prediction. To tackle this issue, we customize the standard mean-squared error loss function by applying additional weights to errors around the junction regions. This allows the model to prioritize learning in high-field areas that significantly impact the calculation of avalanche triggering probability, thereby improving the overall accuracy. The details of the modified loss function, including automatic peak-field region identification and the associated hyperparameters, are discussed in Methods. An example comparison between the predicted and physics-based electric field data is shown in Fig. \ref{fig:fig2_forward_network_illustration}c, where an excellent agreement throughout the nanowire axis can be seen, with the peak-field regions around $x=1000$ and 2000 nm fully captured by the network. The corresponding learning curves with the modified loss function are shown in Fig. \ref{fig:fig2_forward_network_illustration}e. After the training is completed, an overall test mean-squared loss of $4.2\times 10^{-4}$ was obtained. The forward prediction network for quantum efficiencies has an output layer also uniformly sampled to 1000 spatial grid points along the 1D nanowire axis from 0 to 3 $\mu$m. The device physics behind the quantum efficiency is much simpler as it does not involve the highly non-linear avalanche impact ionization in the drift-diffusion model (see Methods). Hence, a typical quantum efficiency profile transitions smoothly from the p-region to the i- and n-regions. The same 80/10/10 data split as in training the electric field network was allocated, employing the mean-squared error as the loss function. A similar example comparison of a set of quantum efficiency data is shown in Fig. 
\ref{fig:fig2_forward_network_illustration}d with the learning performance shown in Fig. \ref{fig:fig2_forward_network_illustration}f. Despite the much smaller training data size, it can be seen that the network captures the physics well and presents very good agreement with the numerical simulation results. 

With the excellent consistency between physics-based simulations and network-predicted electric field and quantum efficiency results, the photon detection efficiency and dark count rate can be readily computed following the fast postprocessing routines (see Fig. \ref{fig:fig1_conventional_workflow}e and Methods). A comprehensive evaluation of our DNN-based surrogate model accuracy is shown in Fig. \ref{fig:fig3_PDE_DCR_mapping_and_FNN_performance}. All 650 sets of simulated ground truth and the forward network prediction are compared for PDE shown in Fig. \ref{fig:fig3_PDE_DCR_mapping_and_FNN_performance}a-c and for DCR shown in Fig. \ref{fig:fig3_PDE_DCR_mapping_and_FNN_performance}d-f, as a function of the i-region length ($L_i$) and the excess bias ($V_e$), which is defined as the difference between the applied voltage and the threshold breakdown voltage. An excess bias is more convenient to use as it always starts from 0 V, regardless of the respective threshold breakdown voltage of a particular case. It can be seen that the 2D mappings of both PDE and DCR predicted by the DNN-based surrogate model greatly resemble the general trend of the physics-based ground truth results. Figs. \ref{fig:fig3_PDE_DCR_mapping_and_FNN_performance}c and f show a more detailed comparison at three representative excess biases ($V_e=0.6$, 2.6, and 4.6 V) along the variation of the i-region length, indicating that the network-predicted results closely follow the ground truth counterparts, in particular, for DCR whose magnitude varies across several orders of magnitude. It is noted that each data point generated from the physics-based simulator takes several hours to compute, whereas the surrogate model only needs milliseconds. 

Overall, the high accuracy of our DNN-based surrogate model proves to be a computationally efficient approximation to the true model, thereby substantially accelerating the candidate design evaluation in a forward design. Although physics-based simulations must still be collected initially to train the networks, this cost is countervailed by the tremendous computational advantages once the network is trained. Then, we incorporate this fully trained surrogate model into our inverse design workflow to form a cascaded network, as will be discussed in the next section.

\subsubsection*{Physics-aware SPAD inverse design workflow}
\noindent
The objective of an SPAD inverse design is to take a target PDE (or DCR) as the input and inform designers of the required region lengths, doping concentrations, and applied voltage, i.e., a vector of [$L_p$, $L_i$, $L_n$, $D_p$, $D_i$, $D_n$, $V_a$]. The central challenge to this inverse problem is that the same PDE and DCR can be potentially achieved by different combinations of nanowire dimensions and doping at a different voltage. Although the distribution of an avalanche electric field along the nanowire, which is the key intermediate quantity for calculating PDE and DCR, is unique for a given combination of lengths, doping, and voltage, such uniqueness is lost after performing a spatial integration to obtain a single-valued PDE. This is quite different from, for example, the metasurface designs where deep-learning-based methods have also been extensively exploited with state-of-the-art performance. In their workflows, the input (e.g., metamaterial geometries) and output (e.g., a desired transmittance for a specific wavelength range) of a forward neural network model can be swapped to train another deep neural network for the inverse designs \cite{An_etal_milestone_2019,Wellie_AEM_DL_Review_AFM2021}. In SPAD inverse designs, although one could potentially train an inverse network by taking an electric field as the input to circumvent the issue of non-unique solutions, it is unrealistic to do so because a detector's electric field can neither be directly measured experimentally nor easily specified by designers like the case in metasurface design by drawing some transmittance spectra with simple step functions. 

Instead, we employ the random decision forests (RF) algorithm, a supervised machine-learning method, to tackle this ``one-to-manyness" issue between a single-valued PDE and many possible options of the vector [$L_p$, $L_i$, $L_n$, $D_p$, $D_i$, $D_n$, $V_a$] as a classification problem, rather than training an inverse neural network and treating it as a regression problem. Generally, deep neural networks trained for inverse problems with non-unique solutions tend to perform poorly since their training algorithms, like gradient descent, implicitly assume the uniqueness of solutions \cite{Willie_InvDesign_Benchmark_Nanoscale_2022}. All 650 sets of physics-based simulation data (Fig. \ref{fig:fig3_PDE_DCR_mapping_and_FNN_performance}a) with calculated PDEs as input and the seven-component vectors as output were used to train the RF model. The RF model is an ensemble learning method that combines the predictions of multiple decision trees to make classifications, in this case, to find a candidate vector of SPAD parameters that best matches an input PDE. Each decision tree in the forest has a flowchart-like structure illustrated in Fig.\ref{fig:fig4_inverse_design_workflow}a, branching from the top, root node (i.e., an input PDE value) to the bottom, leaf nodes (i.e., a prediction of device parameter vector), pass through the internal nodes that split the predictor space (i.e., the set of possible PDE values) based on certain threshold PDE values. Building a decision tree involves optimizing these threshold PDEs assigned to the split nodes such that the variance between the training seven-component vector generated from the leaf node and the target vector is minimized. We used the standard mean-squared error to measure the quality of a split and adopted a minimum of two samples required to split an internal node, with a minimum of one sample required to be at a leaf node. A single decision tree, however, can be non-robust and generally have limited predictive accuracy \cite{james2023introduction}. The key advantage of using an RF model is that by aggregating many decision trees, the predictive performance can be substantially improved, and over-fitting can be avoided \cite{Breiman2001}. Each decision tree independently learns from a bootstrapped sub-sample of the training data, and the RF model then uses an ensemble method called ``majority voting" to aggregate the predictions from all trees. In this process, each tree casts a vote for the potential SPAD parameter vector it predicts, and the vector that receives the most votes across all trees is selected as the final candidate vector for a given PDE. Growing each tree on a bootstrapped resample of the original training data greatly increases the diversity among the ensemble members, and thus the RF model can better learn the complicated inverse relations between the PDE and the nanowire region lengths, doping and applied voltage. Moreover, the RF model offers significantly faster training with fewer hyperparameters to tune. In our case, we constructed the RF model with 500 decision trees, at which point the test error has settled sufficiently low. The 80/10/10 data split for training, testing and validation were used. The training of the RF model only took a few tens of seconds on an Intel Core i7-11370H 3.3 GHz and 8 GB memory. 

The candidate vector generated by the RF model, however, may still not yield exactly the target PDE. We thus connect the RF model to the fully trained forward network models for further evaluation, as shown in Fig. \ref{fig:fig4_inverse_design_workflow}b. The full vector and its first six components are given to the networks of electric field and quantum efficiency, respectively, with all the trained weights and biases fixed. Following the same postprocessing routines discussed in the last subsection, a predicted PDE (PDE$_{\text{predicted}}$) can be immediately calculated and compared against the target PDE (PDE$_{\text{client}}$) within a user-specified error margin ($\epsilon$). If PDE$_{\text{predicted}}$ does not agree with PDE$_{\text{client}}$, we adjust one of the vector components, the applied voltage ($V_a$), according to whether PDE$_{\text{predicted}}$ is greater or smaller than PDE$_{\text{client}}$. In principle, changing region lengths, doping, applied voltage, or a combination of any two or all could alter the resultant avalanche electric field and quantum efficiency, thus changing the PDE value. Based on the device physics, however, increasing (decreasing) the applied voltage is the most straightforward way of enhancing (suppressing) the strength of an avalanche electric field, thereby increasing (decreasing) the PDE. The adjusted vector is fed into the forward models to go through another round of evaluation. The iteration continues until the predicted PDE matches the target PDE.  

Furthermore, DCR is provided as extra information based on the calculated predicted PDE, as shown in Fig. \ref{fig:fig4_inverse_design_workflow}c. Generally, several dark carrier generation mechanisms contribute to DCR \cite{Donnelly_SPAD_model_IEEE_2006,Jiang_SPAD_model_IEEE_2007}. In this work, we include contributions from thermal generation, trap-assisted tunneling, and direct band-to-band tunneling, whose modeling details can be found in Methods. DCR is proportional to the material's bulk defect density, which is an extra input that is not specified in the device parameter vector but can be easily incorporated into the post-processing routine. As stated before, the defect density is not considered a part of the design parameter vector since it is more of an outcome of different combinations of a device's dimensions and doping. By specifying a typical range of defect densities, the corresponding DCR range can be readily calculated, as indicated by the colored double arrows in Fig. \ref{fig:fig4_inverse_design_workflow}c. A large electric field leads to a high avalanche triggering probability for both light-generated and dark carriers, which gives a good PDE but also a large DCR. Usually, designers specify a target PDE with a tolerance range of DCR based on a particular SPAD application. If the requirement of PDE can be relaxed to some extent, the DCR performance can also be improved. In the example shown in Fig. \ref{fig:fig4_inverse_design_workflow}c, such ``relaxation" on PDE is quantified as 5\%, but it can also be specified by designers. Our workflow thus informs designers of a downgraded performance of $\text{PDE}_{\text{client}}-5\%$ with an improved DCR range, as well as a more extreme case of $\text{PDE}_{\text{client}}+5\%$ with a worsened DCR performance. In this process, the values of $\text{PDE}_{\text{client}}\pm5\%$ are calculated via the same iteration loop shown in Fig. \ref{fig:fig4_inverse_design_workflow}b by adjusting the applied voltage. 

Overall, our proposed inverse design workflow does not rely on the time-consuming physics-based simulations but only utilizes fast machine-learning-based methods such as the random forest model and fully trained DNN-based surrogate models, which can be generalized to device structures beyond nanowires. Still, the inverse design delivers high accuracy comparable to its physics-based counterpart, and the workflow's optimization sequence takes a negligible amount of time compared to the iteration cycles in a conventional design workflow, which will be demonstrated by a concrete example in the next subsection.

\subsubsection*{Demonstrating on-demand nanowire SPAD design}
\noindent
Following the same setup of an InP nanowire array-based SPAD introduced in the previous section, we now demonstrate the design of nanowire SPAD employing the proposed inverse design workflow. The overall idea is that designers can specify a target photon detection efficiency (PDE) as an input, and the inverse design workflow will infer the required key nanowire parameters and the associated applied voltage, summarized as a seven-component vector, through its internal optimization loops enabled by the fast machine learning algorithms. To verify the inverse design's accuracy, the inferred nanowire parameter vector is also fed into the physics-based simulator to compute the ground-truth values for comparison. We investigate both the mean PDE and peak PDE as the input for the inverse design workflow. The detailed definition of these two quantities can be found in Methods. In short, the original, raw PDE data is a spatial variable (i.e., PDE($x$)) along the nanowire axis, and its value typically peaks within the high-field region (i.e., the i-region) and decreases quickly as it goes into the adjacent p- and n-regions. The peak PDE is simply the maximal value of a raw PDE($x$) distribution, which is the best performance that can be achieved, provided that extra means, such as light-trapping, are incorporated into the nanowire array design. The peak PDE of a nanowire array-based SPAD can be much higher than its conventional planar structure-based counterparts, which is one of the key promises of employing nanowire-based SPADs. The mean PDE is a full-width at half maximum length-averaged quantity, serving as a lower bound of the corresponding peak PDE value. Both mean and peak PDEs are single-valued quantities that are more convenient to use and cite for performance comparison among different device designs. It is important to verify both mean and peak PDE predictive performance as they reveal different aspects of the proposed inverse workflow. Since the peak PDE is directly extracted from a raw PDE($x$) distribution, it is more sensitive to the locations within a device than spatially averaged mean PDE and, therefore, has a more stringent requirement for the prediction accuracy of the inverse design workflow.

Fig. \ref{fig:fig5_inverse_design_showcase} summarizes the predictive performance of the inverse design workflow, where the vertical dashed lines indicate a range of target mean (Fig. \ref{fig:fig5_inverse_design_showcase}a) and peak (Fig. \ref{fig:fig5_inverse_design_showcase}b) PDEs specified by the designers. An assigned PDE value is first fed into the trained random forest (RF) model to generate a candidate nanowire parameter vector containing the region lengths, doping concentrations, and applied voltage; the components of the candidate vector, particularly the applied voltage, are refined through the device physics guided iteration loop employing the fully trained DNN-based surrogate models until the predicted PDE closely matches the target PDE. These are the blank circle data points in Fig. \ref{fig:fig5_inverse_design_showcase}, denoted as inverse-design generated, where we show a range of representative PDE values from 0.2 to 0.55 for mean PDEs and from 0.2 to 0.7 for peak PDEs. For visual clarity, note that the seven components of these nanowire parameter vectors behind each circular data point were not explicitly shown in the figure. The refined nanowire parameter vector can also be directly fed into the physics-based simulator for verification, giving blank squared data points in Fig. \ref{fig:fig5_inverse_design_showcase}. Each data pair generated from the inverse design and the physics-based simulator is connected by a dashed line to highlight that they are initialized from the same target PDE. The DCR across a broad range of defect densities, from $10^{10}$ cm$^{-3}$ to 10$^{18}$ cm$^{-3}$, is computed and outputted as secondary information for each target PDE. Note that the range of the bulk defect density may not resemble an experimentally realistic range but merely show the flexibility of our inverse workflow. The designers can always adjust the underpinning dark carrier generation models for a specific material system, including the defect density. 

For the predictive performance of mean PDE values shown in Fig. \ref{fig:fig5_inverse_design_showcase}a, it can be seen that all the inverse designs presented by the circle data closely follow the target values. However, for target PDEs less than 35\%, the physics-based results show some deviations. For larger target PDEs, both physics-based verification and inverse-design generated data points closely resemble each other, indicating excellent predictive accuracy of our inverse design workflow. Since the inverse workflow incorporates two distinct machine learning models, i.e., a random forest (RF) classification model and two DNN-based regression models, it is hard to quantify the detailed error contribution of these two modules for small target PDEs. It is likely that due to the relatively small training data size for low PDEs, the candidate nanowire vector generated from the RF classification may have some moderate deviation from a potential ground-truth vector in terms of the region lengths and doping. However, such discrepancy is `magnified' in the subsequent iteration loops that only refine the candidate vector by adjusting the applied voltage, resulting in overfitting and further deviation. For the inverse design using peak PDEs as input, it can be seen that the deviations between inverse designs and physics-based verification become slightly larger. This indicates that the strongly location-dependent peak PDEs are more prone to overfitting, as it requires a more accurate prediction of the peak avalanche electric field that gives rise to the peak PDE value. The deviations in DCR between the inverse designs and physics-based simulations are, however, much smaller, even in cases where PDE values show relatively large discrepancies. This indicates that DCR is much less sensitive to the spatial variations in the avalanche electric field and has a higher tolerance for the prediction errors generated either from the RF model or the DNN-based surrogate models. The reasons are two-fold: 1) like mean PDE, DCR is also a spatially integrated quantity, and 2) the various dark carrier generation rates quickly drop down to a negligible level as the avalanche field falls away from its peak value, making the resultant DCR insensitive to the exact location of the peak field.

\subsection*{Discussion}\label{sec:discussion}
\noindent
Our proposed inverse design workflow for single-photon avalanche detectors features several noteworthy contributions. First, we significantly accelerate the forward device modeling by employing a deep neural network-based surrogate model, replacing the most time-consuming time-dependent drift-diffusion model in a conventional forward design and serving as the building block for the proposed inverse design. The DNN-based surrogate model takes commonly used, experimentally accessible device parameters as input and generates a highly accurate electric field distribution, a key intermediate variable for calculating SPAD's photon detection efficiency (PDE) and dark count rate (DCR). The computation time is reduced by several orders of magnitude from hours to milliseconds, making the evaluation of candidate designs much faster than via conventional drift-diffusion simulations. To the best of our knowledge, this is the first report on applying artificial neural networks to predict electrical characteristics common to semiconductor optoelectronic devices. Second, we realize the on-demand nanowire SPAD design in a few seconds by connecting the trained forward surrogate models to a random forest machine-learning model, taking a preset performance target, i.e., a photon detection efficiency in the demonstration, and generating an optimal design that provides the closest match to the target. This is accomplished without involving numerous computationally intensive drift-diffusion simulations conducted usually in a trial-and-error manner. The accuracy of inferred design parameters can be verified by feeding into the physics-based simulator, where it shows that the average error between our inverse model and ground-truth simulations is less than 10\% for both mean and peak PDEs. Third, the methodology of our inverse design workflow is agnostic of material systems and device architectures (being nanowire- or planar-based structures), making it readily generalizable to other optoelectronic devices. This is because the key input parameters we identified for the forward surrogate model, i.e., the region lengths, doping, and applied bias, which are also the outputs of the inverse design, are universally crucial for the current SPAD technologies. Furthermore, the methodology discussed here is also agnostic of the characteristic, so while we focused on the photon detection efficiency and dark count rate, the approach could be extended to other figures of merit. For instance, the forward surrogate models can be immediately applied to compute afterpulsing probability, which also is a critical metric characterizing an SPAD's internal noise and depending on the distribution of an electric field.

Like all the deep learning-based approaches employed in nanophotonic device designs, the performance and accuracy of the forward surrogate models (and thus the on-demand inverse design) are limited by the quantity and quality of training data. Both the electric-field and quantum-efficiency networks are data-driven models that become increasingly accurate as more training data are fed. So far, the training data generation process, especially for the avalanche electric field, takes much longer than network construction and training. This bottleneck, although greatly countervailed by the computational efficiency once trained, can be alleviated by future open-source data sharing among the research community. Our work serves as an important starting point and will continually contribute to building more comprehensive training data sets. Moreover, another potential resolution to the issue of comparatively little training data is to adopt a transfer learning approach, which is popular in the broader deep-learning community and refers to the reuse of a pre-trained model on a different but related problem. In the context of designing SPADs, this might involve training an electric field prediction sub-network using the model parameters obtained from training a metasurface device, whose input is device geometries similar to the SPAD input, and the output is some electromagnetic scattering response as a function of wavelengths. Based on a range of published work, it suggests that many metasurface forward models present great similarities in the network structures compared to our electric field forward model (e.g., \cite{An_etal_milestone_2019}). Rather than randomly choosing weights and often demanding a large training data size, we could adopt a pre-trained metasurface model as a good initial guess for SPAD model training with relatively small data size while still maintaining high accuracy. 

Lastly, our current inverse design approach deals with the one-to-many mapping issue between a target response and an inferred design by using a random forest classification model. Although the random forest model always produces a single solution due to its majority voting, searching for the closest match to a target response ultimately depends on how densely the training data are packed in the device parameter space. In other words, such an approach is expected to scale poorly with increasing SPAD design complexity. Recent progress in nanophotonic inverse design may again shine some light on this issue. Several recent works have demonstrated that the one-to-many mapping from target to response in metamaterial inverse designs can be directly learned through conditional generative models, such as conditional generative adversarial networks \cite{So2019DesigningNS,Wen2020,An_etal_CGAN_2021}, conditional variational autoencoders \cite{Ma_etal_cVAEs_AM_2021,Lei_etal_cVAEs_2024}, and conditional diffusion models \cite{LEW_diffusion_model_2023,Zhang_etal_diffusion_model_2023,Bastek_etal_diffusion_model_2023}, which will be a future direction of our research to further improving the performance of the SPAD inverse design.

\subsection*{Methods}\label{sec:methods}
\subsubsection*{Training data generation}
\noindent
The physics-based simulations were performed using the Semiconductor Module of COMSOL Multiphysics to generate high-fidelity data to train the forward network surrogate models for electric field and quantum efficiency, respectively. The semiconductor drift-diffusion model (Fig. \ref{fig:fig1_conventional_workflow}b) is a set of three coupled partial differential equations that model the optoelectronic physics of a single-photon detector \cite{Spinelli_SPAD_Physics_and_Modeling_IEEE1997}. It involves solving three variables, i.e., electric potential ($\phi$) and electron ($n$) and hole ($p$) concentrations. To obtain physically correct avalanche electric field distributions with the space-charge effect, the original, time-dependent drift-diffusion model shown below is solved \cite{Sze_PhysSemiDevices_textbook_ch1}:
\begin{align}
    \triangledown \cdot (\varepsilon \phi) &= -q (p-n + N_D^+ - N_A^-), \\
    \frac{\partial n}{\partial t} &= G_n^{\text{impact}} -R_n + \frac{1}{q} \triangledown \cdot \boldsymbol{J}_n, \\
    \frac{\partial p}{\partial t} &= G_p^{\text{impact}} -R_p - \frac{1}{q} \triangledown \cdot \boldsymbol{J}_p,
\end{align}
where $q$ is the elementary charge, $\varepsilon$ is the dielectric constant of the semiconductor material, and $N_D^+$ and $N_A^-$ are ionized donor and acceptor concentrations, respectively. $G_n^{\text{impact}}$ and $G_p^{\text{impact}}$ are the electron and hole generation rates, respectively, due to the impact ionization that can trigger an avalanche event. They are a function of both the local electric field and the material-related parameters called impact ionization coefficients ($\alpha_n$ and $\alpha_p$). We adopt the ionization coefficient model based on Monte Carlo simulations, verified by experimental data for temperatures ranging from 150 to 290 K. $R_n$ and $R_p$ are the electron and hole recombination rates modeled by the standard Shockley-Read-Hall model \cite{Sze_PhysSemiDevices_textbook_ch1}, which are a function of the carrier lifetime. For the demonstration cases where InP nanowires are considered, 1 ns was used based on the experiments \cite{QianGao_InP_NW_review_2019}. The electron ($\boldsymbol{J}_n$) and hole ($\boldsymbol{J}_p$) current density are given by the following constitutive relations:
\begin{align}
    \boldsymbol{J}_n &= -q n \mu_n \triangledown \phi + q D_n \triangledown n, \\
    \boldsymbol{J}_p &= -q p \mu_p \triangledown \phi - q D_p \triangledown p,
\end{align}
where $\mu_n$ and $\mu_p$ are the electron and hole mobilities, respectively, and $D_n$ and $D_p$ are the electron and hole diffusion coefficients, respectively. Solving the time-dependent drift-diffusion equations above gives a time-dependent, spatial electric potential distribution $\phi(x,t)$. For data training, we only extract its steady-state distribution $\bar{\phi}(x)$, which is determined by monitoring the gradient of the detector's terminal current with respect to time to be less than a preset threshold value \cite{Li_SPAD_workflow_2023}. The electric field $\boldsymbol{E}$ can be derived through the fundamental relation $\boldsymbol{E} = - \triangledown \bar{\phi}$.

The quantum efficiency is calculated by solving the steady-state drift-diffusion model:
\begin{align}
    \triangledown \cdot (\varepsilon \phi) &= -q (p-n + N_D^+ - N_A^-), \\
    0 &= G_n^{\text{light}} -R_n + \frac{1}{q} \triangledown \cdot \boldsymbol{J}_n, \\
    0 &= G_p^{\text{light}} -R_p - \frac{1}{q} \triangledown \cdot \boldsymbol{J}_p,
\end{align}
where the time dependence on the left-hand side of the electron and hole continuity equations are approximated as zero. The impact ionization terms are also replaced with the light carrier generations since the quantum efficiency characterizes intrinsically how well a detector absorbs photons and collects the associated electron-hole pair, regardless of whether there is an avalanche. Removing the highly non-linear impact ionization terms also makes the drift-diffusion equations much easier to solve. The spatial quantum efficiency distribution $\text{QE}(x)$ can be calculated as:
\begin{equation}
    \text{QE}(x) =  \frac{I_{\text{light}}(x)/q}{G^{\text{light}} \mathcal{V}},
\end{equation}
where $G^{\text{light}}=G_n^{\text{light}}=G_p^{\text{light}}$, and $I_{\text{light}}$ is the photocurrent measured from the detector's terminal, whose magnitude depends on which part of a nanowire is optically excited (thus a function of $x$). In principle, the choice of the absolute value of a $G^{\text{light}}$ is less important in determining $\text{QE}(x)$ as the corresponding photocurrent $I_{\text{light}}$ scales with $G^{\text{light}}$. $\mathcal{V}$ is the volume where $G^{\text{light}}$ is applied, which can be approximated as $\mathcal{V}= A d$, where $A$ is the nanowire cross-sectional area and $d$ is the optical excitation size (i.e., averaged beam size). Here, an experimentally feasible nanowire diameter of 150 nm was used \cite{QianGao_InP_NW_review_2019}, which gives a cross-section area of 0.018 $\mu$m$^2$. The optical excitation size $d$ can be set to a realistic value based on the laser spot size, or it can be smaller if a detailed mapping of quantum efficiency is desired. In this work, an excitation size $d=50$ nm was used to obtain a detailed mapping that characterizes the NW intrinsic photoresponse. Initially, the excitation ‘box’ is placed at the p-doped end of NW and is progressively scanned through the entire NW with a step size of 50 nm. For each location where the optical excitation is applied, the steady-state drift-diffusion simulation was performed to obtain one data point of the quantum efficiency at that location. In total, it generates a spatial quantum efficiency distribution of 60 grid points, which is subsequently interpolated to 1000 grid points for the ease of forward network training.

\subsubsection*{Calculation of dark count rate}
\noindent
The calculation of the dark count rate (DCR) relies on a key intermediate quantity called avalanche triggering probability, which uses the electric field as an input. Avalanche triggering probability characterizes the probability of the generated electron-hole pair successfully triggering a self-sustaining avalanche. We denote avalanche triggering probability as $P_{\text{pair}}$ thereafter, which is defined as:
\begin{equation}
    P_{\text{pair}}(x) = P_e(x) + P_h(x) - P_e(x) P_h(x),
\end{equation}
where $P_e$ and $P_h$ are the probabilities that an electron or a hole can trigger an avalanche event, respectively. Once $P_e$ and $P_h$ are known, $P_{\text{pair}}$ can be readily calculated. Generally, for a p-i-n structure, either in a nanowire or a planar bulk material, $P_e$ and $P_h$ can be solved via the following differential equations \cite{McIntyre_1973_ATP}:
\begin{align}
    \frac{\partial P_e}{\partial x} &= -(1-P_e) \alpha_e (P_e+P_h-P_e P_h), \\
    \frac{\partial P_h}{\partial x} &= (1-P_h) \alpha_h (P_e+P_h-P_e P_h),
\end{align}
where $\alpha_n$ and $\alpha_p$ are the electron and hole impact ionization coefficients, respectively, whose formulations and relevant InP-related parameters can be found in \cite{Petticrew_2020_modeling_II_InP}. With the solved $P_{\text{pair}}$, DCR can be determined by: 
\begin{equation}
    \mbox{DCR} = A \int_0^L P_{\text{pair}}(x) G_{\text{tot}}(x) dx,
\end{equation}
where $A$ is the cross-sectional area of a nanowire, $L$ is the length of a nanowire, and $G_{\text{tot}}$ is the total dark carrier generation rate. In this work, we include contributions from thermal generation based on the Shockley-Read-Hall model, direct band-to-band tunneling, and trap-assisted tunneling, whose detailed formulations for InP nanowire can be found in \cite{Li_SPAD_workflow_2023}. 

\subsubsection*{Calculation of photon detection efficiency}
\noindent
The calculation of the photon detection efficiency (PDE) takes the avalanche triggering probability and quantum efficiency as the input and is given by:
\begin{equation}
    \text{PDE}(x) = \text{QE}(x) \times P_{\text{pair}}(x),
\end{equation}
where the raw data of the photon detection efficiency is a spatial dependent variable. The peak PDE demonstrated in Fig. \ref{fig:fig5_inverse_design_showcase}b is the maximum value directly extracted from this spatial distribution. The mean PDE, which is defined as a full-width at half-maximum length-averaged PDE$(x)$, is given by:
\begin{equation}\label{eq:length_avg_PDE}
    \langle \text{PDE} \rangle = \frac{1}{x_2-x_1} \int_{x_1}^{x_2} \text{PDE}(x) dx,
\end{equation}
where $x_1$ and $x_2$ define the boundary $x$ coordinates that cover the half maximum. The mean PDE is thus a single-valued quantity for a given combination of nanowire dimension, doping concentration, and applied voltage. 

\subsubsection*{Forward network constructions}
\begin{figure} 
	\centering
	\includegraphics[width=1.0\textwidth]{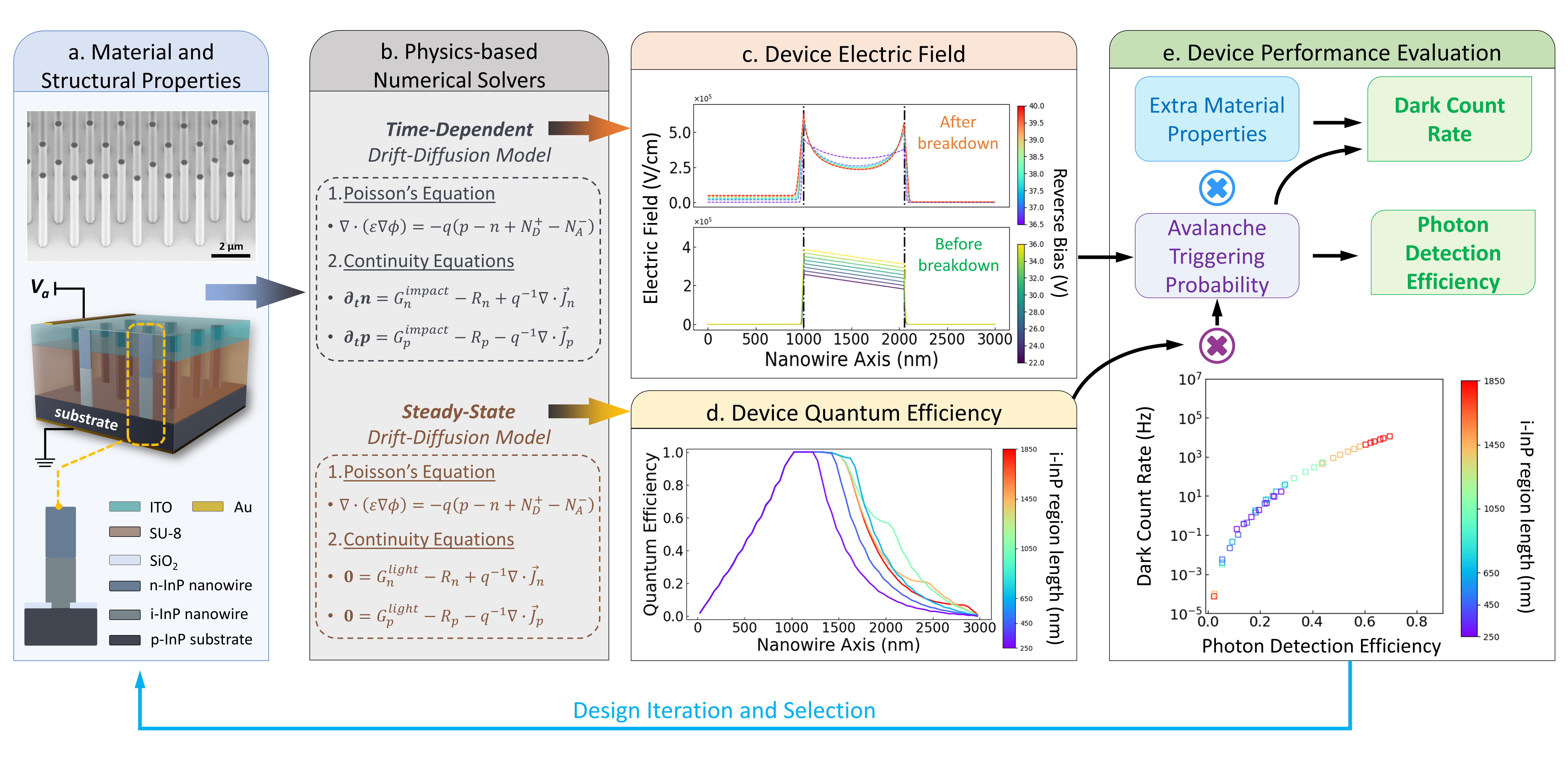} 

	\caption{\textbf{A typical conventional single-photon avalanche detector (SPAD) design workflow.} a, Material and device structure selection. An indium phosphide (InP) nanowire-based SPAD is used as an example. A scanning electron microscope (SEM) image of an as-grown InP nanowire array is shown in the top panel. A device schematic is illustrated in the middle panel, with an enlarged view of a unit cell shown at the bottom. b, Physics-based models for simulating the intermediate quantities critical to computing the key SPAD performance metrics. c, Electric fields simulated by the time-dependent drift-diffusion model at external biases below the device's breakdown voltage (lower panel) and above (upper panel). d, Quantum efficiencies simulated by the steady-state drift-diffusion model as a function of the intrinsic region length. e, A common scheme of computing the photon detection efficiency (PDE) and dark count rate (DCR) based on the intermediate quantities obtained in the previous steps. A typical relation between DCR and PDE for different intrinsic region lengths is also illustrated in the lower panel.}
	\label{fig:fig1_conventional_workflow} 
\end{figure}

\begin{figure} 
	\centering
	\includegraphics[width=1.0\textwidth]{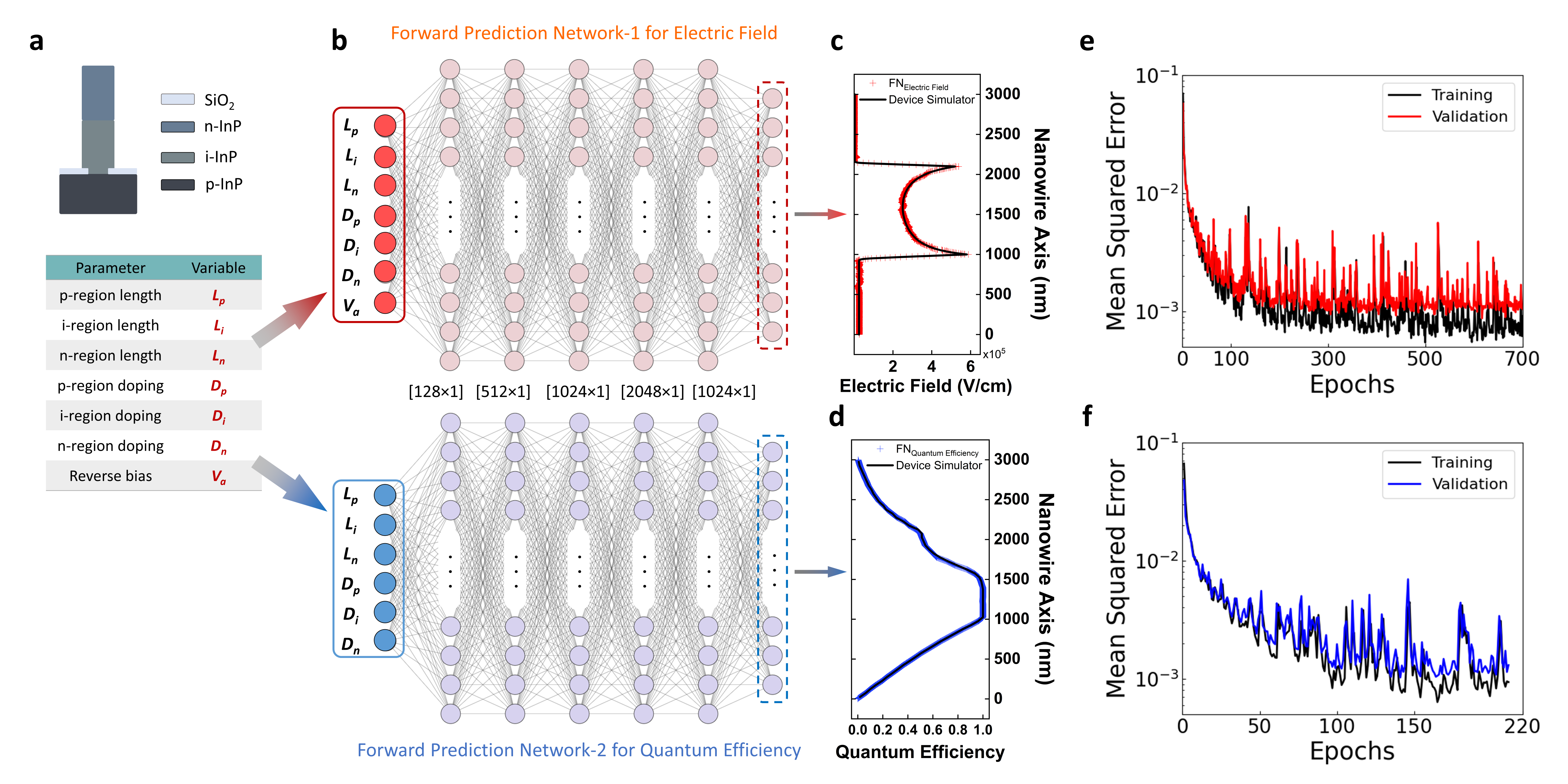} 

	\caption{\textbf{Method and structure of the two forward prediction networks.} a, A unit-cell device schematic extracted from a nanowire array. Seven parameters of the unit cell are identified as the key input for the neural networks, summarized in the table at the bottom. b, the structures of the two forward prediction neural networks, one for electric fields (upper panel), and the other for quantum efficiency (lower panel). The forward prediction network for electric fields takes all seven parameters as its input and generates a spatial distribution of an electric field along the one-dimensional axis of the nanowire unit cell. The forward prediction network for quantum efficiencies does not need the reverse bias ($V_a$) as its input and also generates a position-dependent quantum efficiency along the nanowire unit cell. c and d, An example comparison between an electric field (or a quantum efficiency distribution) generated from the forward network and the physics-based device simulator at a certain set of input parameters. e and f, The learning curves of the two forward prediction networks.}
	\label{fig:fig2_forward_network_illustration} 
\end{figure}

\begin{figure} 
	\centering
	\includegraphics[width=1.0\textwidth]{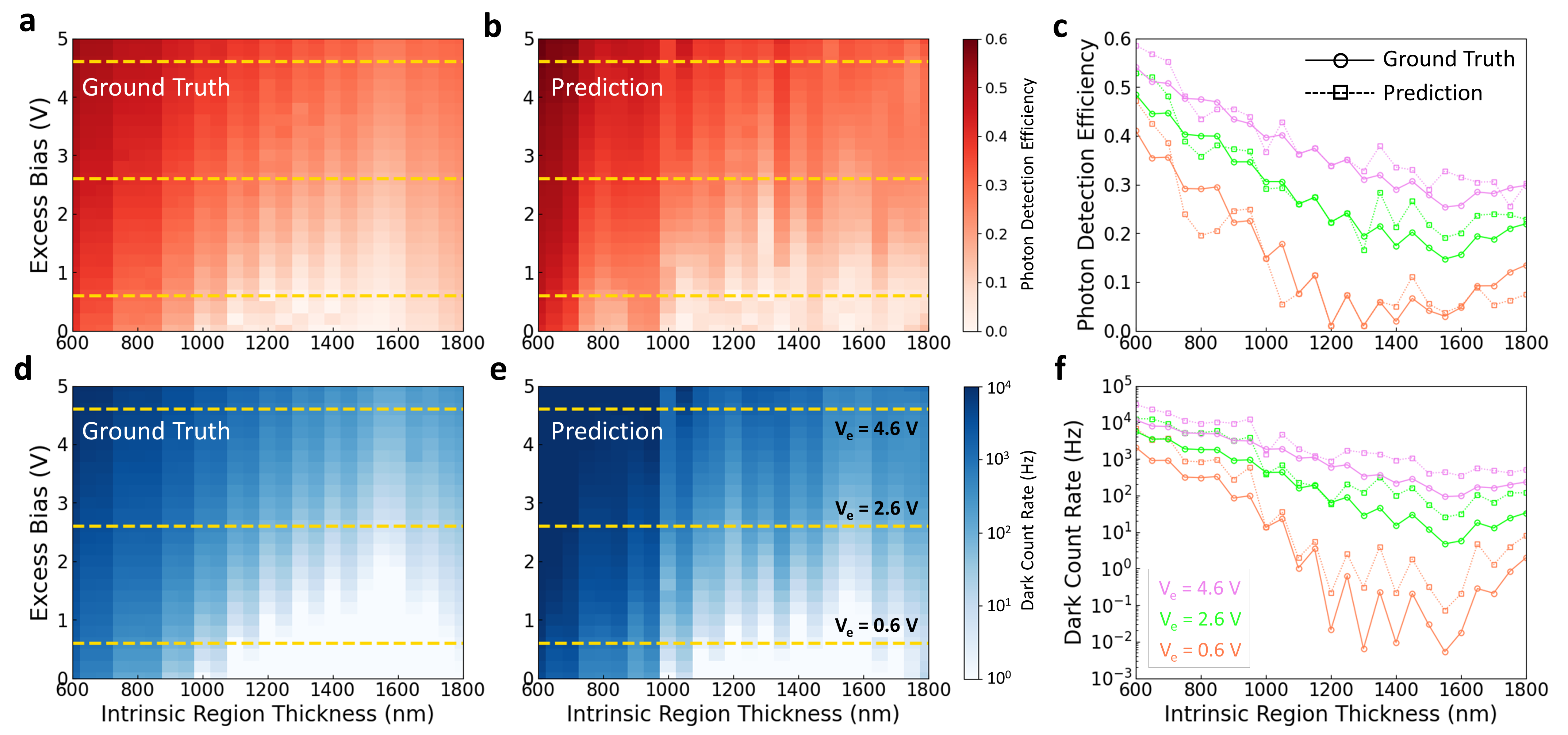} 

	\caption{\textbf{The comparison of photon detection efficiency and dark count rate generated by the physics-based simulations and the prediction networks.} The photon detection efficiency and dark count rate computed by the physics-based simulation results are shown in (a, d), respectively, and those generated from the forward prediction networks are shown in (b, e), respectively. The physics-based training data were computed for the intrinsic region lengths varied from 600 to 1800 nm with a step size of 50 nm, each of which was biased at excess voltages ranging from 0 to 5 V beyond the avalanche breakdown with a step size of 0.2 V. A zoomed-in comparison of PDE and DCR extracted at three excess biases of 0.6, 2.6, and 4.6 V is shown in (c, f), respectively. The locations of these three excess bias values are also highlighted in the golden dashed lines in the 2D heatmap plots.}
	\label{fig:fig3_PDE_DCR_mapping_and_FNN_performance} 
\end{figure}

\begin{figure} 
	\centering
	\includegraphics[width=1.0\textwidth]{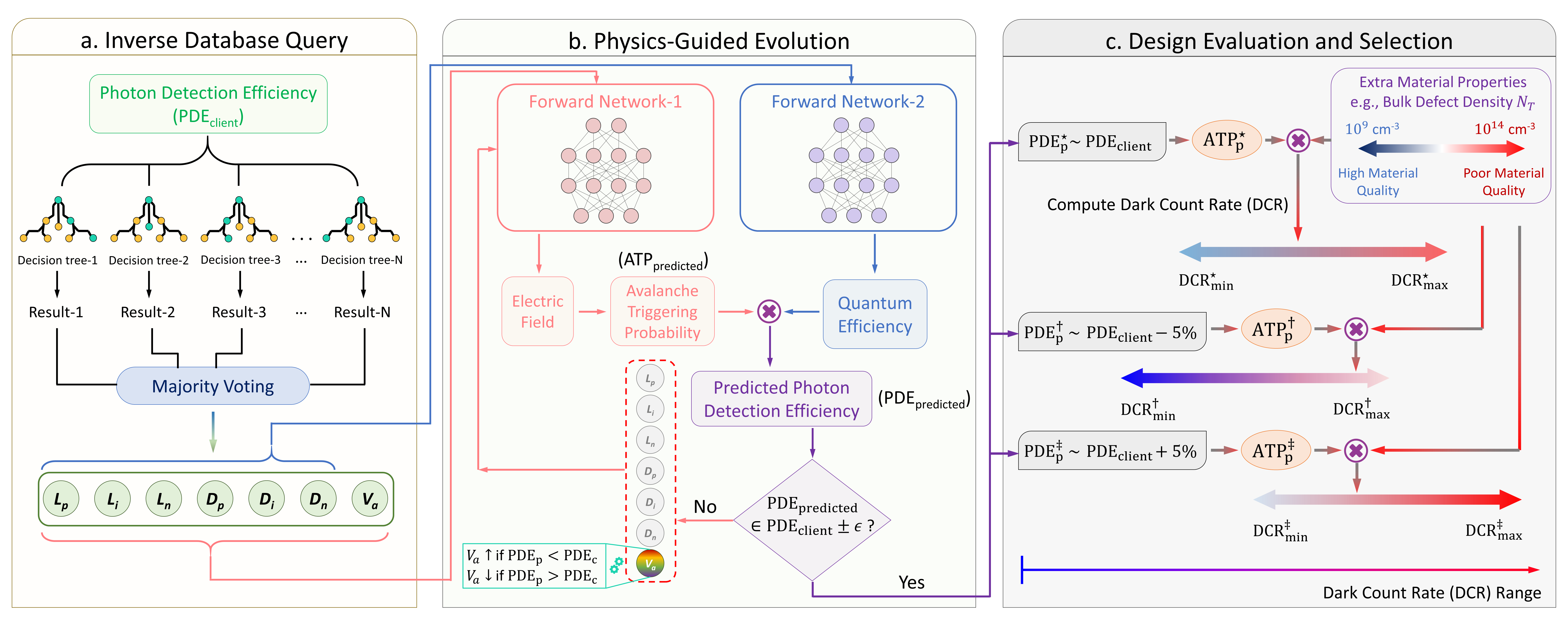} 

	\caption{\textbf{The objective-driven inverse design workflow for single-photon avalanche detectors.} a, The inverse query workflow. The photon detection efficiency (PDE) is chosen as the single objective of the inverse design workflow. A client's photon detection efficiency (PDE$_{\text{client}}$) is designated as the input to the random forest algorithm, which generates a single output vector containing all the nanowire structural and material properties at the required excess bias that can achieve the client-specified PDE. b, The output vector from step (a) is fetched to the two trained forward networks to compute the corresponding electric field and quantum efficiency, where the former can be used to calculate a predicted avalanche triggering probability (ATP$_{\text{predicted}}$). The quantum efficiency and ATP$_{\text{predicted}}$ are then used to obtain a predicted photon detection efficiency (PDE$_{\text{predicted}}$). The predicted PDE and client-specified PDE are compared within a user-set margin $\epsilon$. If the two do not agree, a new input vector with the adjusted excess bias is fed into the forward network for electric fields to obtain an updated electric field. This regulation process is guided by the SPAD device physics as PDE generally increases with an increasing excess bias. c, Once PDE$_{\text{predicted}}$ agrees with PDE$_{\text{client}}$ within a margin, the corresponding dark count rates (DCRs) can be readily calculated for a range of material defect densities. Our workflow also computes DCRs for a broader range of client-specified PDE. If the requirement for a PDE can be relaxed by 5\%, as an example shown in the middle panel, the corresponding DCR can also be lower. The range of the percentage is adjustable and dependent on specific applications.}
	\label{fig:fig4_inverse_design_workflow} 
\end{figure}

\begin{figure} 
	\centering
	\includegraphics[width=1.0\textwidth]{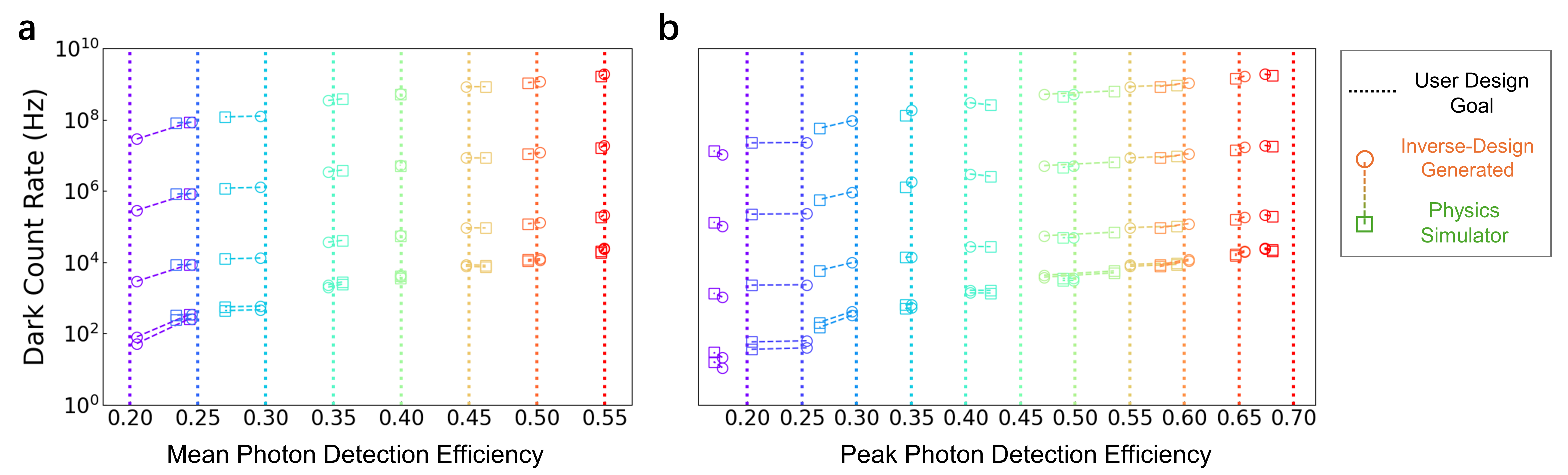} 

	\caption{\textbf{Design examples of the client-specified photon detection efficiency.} a, Mean photon detection efficiency. b, Peak photon detection efficiency. The vertical dashed lines indicate the target PDEs to be fetched into our inverse design workflow. The circle data in the same color are generated from the inverse design. As a benchmark test, the square data are generated from physics-based simulation using the same input vector obtained from the inverse workflow and contains the information on the nanowire's structural, material and electrical properties. The data points generated from the inverse design and physics simulator are connected by a dashed line to highlight they originate from the same target PDE for better visual clarity. The PDEs are plotted for a range of DCRs, using the defect densities of $10^{10}$, $10^{12}$, $10^{14}$, $10^{16}$, and $10^{18}$ cm$^{-3}$.}
	\label{fig:fig5_inverse_design_showcase} 
\end{figure}

\noindent
We constructed two independent forward neural networks to calculate avalanche electric field and quantum efficiency, respectively. All the neural networks in this work were implemented with the open-source machine-learning framework TensorFlow. Data pre-/post-processing and pipelined PDE and DCR computations were implemented with the open-source Python libraries SciPy and NumPy. A total of 1300 configurations of nanowire region lengths and applied voltage, as well as their corresponding avalanche electric field profiles, were collected and used for the training and testing of the electric field forward neural network. Another set of 50 groups of nanowire region lengths and the resultant quantum efficiency data were used to train the quantum efficiency neural network. Both the electric field and quantum efficiency networks consist of five fully connected hidden layers, made of 128, 512, 1024, 2048, and 1024 nodes, respectively, as shown in Fig. \ref{fig:fig2_forward_network_illustration}b. The input layer of the electric field network has seven nodes, representing the nanowire p-, i-, and n-region length, nanowire p-, i- and n-region doping concentration, as well as the applied voltage. The input layer of the quantum efficiency network takes the first six nodes of the electric field network's input layer. The output layer of the two networks is 1000 uniformly sampled spatial coordinates along a nanowire axis. The rectified linear unit (ReLU) function was used as an activation function between the layers. A mini-batch gradient descent learning was implemented during the training, with batch sizes of 32 and 8 for the electric field and quantum efficiency, respectively. The training was performed with the maximum epoch numbers of 1000 (electric field) and 250 (quantum efficiency) until the early stopping point was reached, where the validation loss no longer showed significant improvement to prevent overfitting. Both networks employed the mean-squared error as a cost function and the ADAM solver as the optimizer. In particular, the cost function for training the electric field was customized to take into account the highly nonlinear nature of its distribution at the p-i and n-i junction interfaces (see Fig. \ref{fig:fig1_conventional_workflow}c). All the other hyperparameters, like the learning rate, were adjusted automatically by the TensorFlow framework. 

A typical avalanche electric field shown in Fig. \ref{fig:fig1_conventional_workflow}c indicates that the field strength at the p-i and n-i junctions is a factor of 2 to 3 higher than that in the bulk regions. A less accurate prediction of these peak values by the network surrogate model can result in a significant error in the calculated photon detection efficiency. Therefore, we customized the mean-squared loss function by dynamically applying larger weights to errors in the high-field regions. More specifically, this involves the following steps:
\begin{enumerate}
    \item The absolute difference in the electric field between two neighboring grid points is calculated as: $\delta_i^{(j)} = |y^{(j)}_{i+1}-y^{(j)}_i|$ for $i=1,2,...,n-1$, where $y^{(j)}_i$ refers to the $i$th grid point of $j$th sample in a batch. Then, for each sample,  a mean difference value $\mu^{(j)}$ can be obtained by $\mu^{(j)}=(n-1)^{-1} \sum_i^{n-1} \delta^{(j)}_i$, where $n=1000$ is the number of the sampled grid points of a nanowire axis.
    \item Based on the calculated mean difference in the electric field, we define a threshold difference to identify the peak electric field regions as: $threshold^{(j)}=LH_1\times \mu^{(j)}$, where $LH_1$ is a hyperparameter for the customized loss function that can filter out the baseline difference in the nanowire's bulk region. $LH_1=1.5$ was used after optimization. The indices of the grid points whose mean difference is greater than the threshold value will be returned for the subsequent heavier loss penalty. Thus, the indices of the peak region $I^{(j)}_{\text{peak}}$ is given by: $I^{(j)}_{\text{peak}}=\{i| \delta^{(j)}_i> threshold^{(j)}\}$. 
    \item For the non-peak regions, a baseline weight of 1.0 is assigned to the loss function, whereas for the peak-field regions, a larger weight is assigned, i.e., $w_k^{(j)}= LH_2$ for $k \in I^{(j)}_{\text{peak}}$. $LH_2$ is another hyperparameter for the customized loss function, which was taken as 80 after optimization. 
    \item The modified mean-squared error for each sample is thus given by: $L^{(j)}=n^{-1} \sum_i^n w_i^{(j)}\times \left( y^{(j)}_i - \hat{y}^{(j)}_i \right)^2 $, where $\hat{y}^{(j)}_i$ is the predicted electric field value at $i$th grid point of the $j$th sample. 
    \item Lastly, the batch-wise aggregation of the individual sample loss into a composite loss is calculated by the simple average: $L_{\text{EF}}=m^{-1}\sum_j^m L^{(j)}$, where $m$ is the number of samples in a batch and `EF' denotes `electric field'. 
\end{enumerate}



\clearpage 

%
\bibliography{science_template,spad_inverse_design-bibliography} 
\bibliographystyle{sciencemag}

%
%
%
%
%
%


\section*{Acknowledgments}
The authors acknowledge the financial support from the Australian Research Council. This research was also undertaken with the assistance of resources and services from the National Computational Infrastructure (NCI), which is supported by the Australian Government.
\paragraph*{Funding:}
Z.L., H.T., C.J., and L.F. acknowledge the financial support from the Australian Research Council Centre of Excellence for Transformative Meta-Optical Systems (CE200100010). D.D. acknowledges the support of an Australian Research Council Future Fellowship through project FT220100656.
\paragraph*{Competing interests:}
The authors declare that they have no competing interests.
\paragraph*{Data and materials availability:}
All data needed to evaluate the conclusions in the paper are present in the paper and/or the Supplementary Materials.

\end{document}